\documentstyle{article}
\newcommand{\bra}[1]{\left \langle #1 \right \vert }
\newcommand{\ket}[1]{\left \vert #1 \right \rangle}

\begin{document}
\title{Quantum entanglement and classical separability in NMR computing}
\author{Alexander R. Kessel and Vladimir L. Ermakov \thanks{e-mail: ermakov@dionis.kfti.knc.ru}\\
{\it Kazan Physico-Technical Institute, }\\
{\it Kazan Science Center, }\\
{\it Russian Academy of Sciences,} \\
{\it Kazan 420029 Russia}}
\date{\today}
\maketitle
\begin{abstract}
In the discussion about the quantumness of NMR computation a conclusion is done that 
computational states are separable and therefore can not be entangled. This conclusion is 
based on the assumption that the initial density matrix of an individual molecule coincides with 
whole sample molecules distribution over single molecule energy levels. This means that 
quantum stochasticity is replaced by classical stochasticity. In the present paper it is shown, 
that quantum NMR computation can create genuine entangled states if initial system states are 
thermodynamical equilibrium ones. A separability analysis problem can arise when one 
interprets the readout signal from whole sample. 
\end{abstract}

PACS No.: 03.65.-w, 03.67.-a, 76.60.-k.


\section{ Introduction}
\par It is commonly believed, that one of the reasons of the high effectiveness of 
quantum computations in comparison with the classical ones lays in the usage of the so-
called entangled states of quantum objects. For example, entangled state of two quantum 
objects

\begin{equation}\label{entangled}
\begin{array}{l}
\ket{\Psi}= a \ket{\psi_1} \ket{\psi_2} + \ket{\psi'_1} \ket{\psi'_2} 
\neq \ket{\psi''_1} \ket{\psi''_2},  \\
|a|^2 + |b|^2 = 1.
\end{array}
\end{equation}

\noindent is constructed from the separate objects wave functions 
$\ket{\psi_1}$ and $\ket{\psi_2}$ in such a way, that 
it can not be represented as a product of wave functions of the objects - that is it is non 
separable. 
\par Recently the works \cite{Zyczkowski_On_the_volume}-\cite{Braunstein_Separability} 
have been published, from which it follows, that the 
density matrices of NMR system lay in the proximity of the maximally mixed state and 
that is why they are separable. As a result the NMR computations should be less effective 
comparing to what was expected before. In the present paper it is shown, that in suitable 
NMR experiments there is an entanglement in full extent and that the separability problem 
can arise during  interpretation of the whole sample readout signal.

\section{The arguments in favor of NMR system states separability}
 
\par When analyzing the separability problem and possible entangled states absence in 
the NMR quantum computing usually the following reasoning is given.
\par In the NMR quantum systems a cluster of a small number $N$ of interacting spins 
(for convenience let us call it ``molecule'') plays a role of quantum processor. This 
molecule does not interact with other molecules of sample. The observed signal is a sum 
of signals from $M$ molecules-``computers'' of a whole sample. In liquid state NMR the 
values $\Delta E$ of typical intervals between molecule energy levels satisfy the condition 
$\Delta E << kT$. Then the energy levels population differences are of the order of 
$M\varepsilon$ (where $\varepsilon = \Delta E / kT << 1$) 
and represent a small part of the total number of molecules $M$. At such 
conditions all of spin energy $K = 2^N$ levels of individual processor are occupied 
approximately equally and spin system initial state is described by the mixed density 
matrix

\begin{equation}\label{rho_definition}
\begin{array}{l}
\rho_m = \sum_{k=1}^{K}P_k \ket{\phi^k_m}\bra{\phi^k_m},
\end{array}
\end{equation}

\noindent where $P_k$ is the independent of $m$ possibility to find a molecule in 
a pure spin state $\ket{\phi^k_m}$. 
A quantum system state is called a maximally mixed state if all probabilities equal 
$P_k = 1/K$. Thus in liquid state NMR experiments the small parameter $\varepsilon$ 
means that during 
computations the system states are in the proximity of the maximally mixed state. 
\par In the paper \cite{Zyczkowski_On_the_volume} a mathematical proof is given that there is a small proximity of the 
maximally mixed state where density matrix is separable, that is it can be represented as a 
sum of products of individual particle density matrices. In the paper \cite{Vidal_Robustness} 
the lower limit of 
this proximity is given, whereas in the paper \cite{Braunstein_Separability} 
its upper limit is given too. These 
mathematical proofs are irreproachable. It means that the NMR quantum computation 
effectiveness is much less than in the case when states are entangled.

\section{The arguments in favor of the NMR system states entanglement}

\par We think that the above mentioned separability analysis does not take into full 
account the underlying physics of the NMR experiments. The matter of fact is that during 
computations each processor carries on calculations independently of the states of other 
molecules. And only at the readout step the individual molecules signals are summed up. 
To interpret the results of quantum mechanical measurements as usually one must take into 
account that one deals here with ``double stochasticity'' - the quantum and the classical 
ones \cite{Klyshko_operational_viewpoint}. 
The quantum one should be used when one averages over quantum object states 
which are defined by quantum wave functions. Whereas the classical one should be used 
when one averages over initial states of molecules. 
\par To adequately describe the NMR experiments one should perform two steps:
\par a) to track down the detailed quantum dynamics of individual ``processor'', 
each time 
starting with a possible initial state (using, for example, the wave function language); 
these steps should be repeated for all possible initial states;
\par b) to average results over all possible initial states of all $M$ molecules-``processors'' 
(using density matrix formalism or an appropriate classical distribution - the Boltzmann 
one, the Maxwell one and so on).
\par Following this plan one can note, that at each time the state of an individual 
molecule in general is given by wave function 

\begin{equation}\label{superposition}
\begin{array}{l}
\ket{\phi_m} = \sum_{k=1}^{K} a(t)^k_m \ket{\chi^k_m},
\end{array}
\end{equation}

\noindent where $a(t)^k_m$ 
- numerical coefficients and $\ket{\chi^k_m}$ is the $k$-th eigenstate of the molecule spin 
energy operator. 
\par It is well established in the NMR experiments fact that in the thermodynamical 
equilibrium a molecule occupies one of its stationary states 

\begin{equation}\label{equilibrium}
\begin{array}{l}
\ket{\phi_m}_{eq} = \ket{\chi^k_m},
\end{array}
\end{equation}

\noindent which in the density matrix formalism can be written as 
$\rho_{m,eq} = \ket{\chi^k_m}\bra{\chi^k_m}$. This fact follows 
also from contemporary decoherence theory \cite{Mensky_Decoherence}, 
which describes how an open quantum system 
interacts with environment. 
According to the theory the coherent state (\ref{superposition}) relaxes to one of the 
stationary state (\ref{equilibrium}).
\par During quantum computation such a molecule state may be transferred to non 
separable, entangled state by a few ways and so this state will be involved in 
computation independently of the states of other ``processors''. These calculations are the 
subject of quantum dynamics and not of statistical mechanics. The necessity in the latter 
arise when one reads out a result, which is a sum of individual molecule signals. 

\section{Signal from whole sample}

\par A sample state is defined by the direct product of individual molecule wave functions

\begin{equation}\label{sample_equilibrium}
\begin{array}{l}
\ket{\Psi}_{eq} = \ket{\phi_1}_{eq} \otimes \ket{\phi_2}_{eq} \ldots 
\otimes \ket{\phi_M}_{eq} = \otimes_{m=1}^{M}\ket{\phi_m}_{eq}   
\end{array}
\end{equation}

\par Suppose that a sample is subjected to a sequence of external influences, which realize 
a quantum algorithm. The corresponding propagator is equal to $U_{comp} = \Pi_i U_i$, 
where $U_i$ is a 
unitary transformation, which realizes a logic gate. And if one of these propagators transforms 
a molecule state to an entangled one, then the other gates will continue to work with such a 
state. 
\par Under the influence of propagator $U_{comp}$ the initial spin system state of the whole 
sample (\ref{sample_equilibrium}) transforms to 

\begin{equation}\label{sample_time}
\begin{array}{l}
\ket{\Psi}^{t} = \ket{\phi_1}^{t} \otimes \ket{\phi_2}^{t} \ldots 
\otimes \ket{\phi_M}^{t} = \otimes_{m=1}^{M}\ket{\phi_m}^{t}   
\end{array}
\end{equation}

\noindent where $\ket{\phi_m}^{t}= U_{comp}\ket{\chi^k_m}$.
\par Suppose that to readout a result one has to measure the expectation value of some 
operator $J^x$. Operators of physical quantities are additive 

\begin{equation}\label{Jx}
\begin{array}{l}
J^x = \sum_{m=1}^{M}I^x_m,
\end{array}
\end{equation}

\noindent and that is why in the basis (\ref{sample_time}) 
the expectation value of operator $J^x$ is 

\begin{equation}\label{Jx_t}
\begin{array}{l}
{}^t \bra{\Psi} J^x \ket{\Psi}^t = 
\sum_{m=1}^{M} {}^t \bra{\phi_m} I^x_m \ket{\phi_m}^t \\ 
=  \sum_{m=1}^{M} \bra{\chi^{k_m}_m}U^{+}_{comp} I^x_m  U_{comp}\ket{\chi^{k_m}_m}.
\end{array}
\end{equation}

\par Since the sample molecules are supposed to be identical, then the right hand side 
matrix elements of Eq. (\ref{Jx_t}) depend only on the initial state index $k_m$ and not 
on the molecule number $m$. 
For all molecules with the same initial state $k$ the matrix elements are the same and 
as a consequence the sum over $m$ in the right side of Eq. (\ref{Jx_t}) 
is reduced to the sum over $k$

\begin{equation}\label{Jx_Ck}
\begin{array}{l}
< J^x > \equiv \sum_{k=1}^{K} C_k \bra{\chi^{k}_{m'}}U^{+}_{comp} I^x_{m'}  
U_{comp}\ket{\chi^{k}_{m'}}.
\end{array}
\end{equation}

\noindent where $C_k$ is the number of molecules which initially have been in 
the state $k$ and $m'$ - the index 
of an arbitrary chosen molecule in state $k$. The energy levels population 
numbers $C_k$ are 
defined by suitable distribution function, for example, by the Boltzmann distribution, and 
satisfy the condition 

\begin{equation}\label{Ck}
\begin{array}{l}
\sum_{k} C_k = M.
\end{array}
\end{equation}

\par The expression (\ref{Jx_Ck}) can be written in the form 

\begin{equation}\label{Jx_Sp}
\begin{array}{l}
<J^x> \equiv M Sp(U^{+}_{comp}\rho_{m'}U_{comp}I^x_{m'})
\end{array}
\end{equation}

\begin{equation}\label{rho}
\begin{array}{l}
\rho_{m'} = (1/M) \sum_{k=1}^{K} C_k \ket{\chi^{k}_{m'}} \bra{\chi^{k}_{m'}}.
\end{array}
\end{equation}

From formal point of view the operator 
(\ref{rho}) can be interpreted as a statistical operator of the 
molecule $m'$, which initially was in the mixed state of the kind of 
(\ref{rho_definition}) with $P_k = C_k /M$. 
This fact 
tempts to analyze the states' separability by considering the individual molecule 
$m'$ with its 
``averaged'' density matrix (\ref{rho}) and multiplying the result 
by the total molecules number $M$. 
However such an operation replaces quantum stochasticity with the classical one. As it was 
pointed out in the introduction, the molecule spin levels populations $C_k$ and therefore the 
probabilities $P_k$ with different $k$ are approximately equal . Then formally density matrix 
(\ref{rho}) 
belongs to the proximity of maximally mixed state. 
The misuse of the {\it statistical} ({\it ``averaged'' } over 
sample) density matrix (\ref{rho}) for description of {\it real} quantum states of an 
{\it individual} molecule 
leads to the conclusion that {\it real individual} wave functions remain separable during 
computation. The important point is that the statistics, used in (\ref{rho}), 
describes the {\it number} of 
molecules on the different energy levels, whereas the {\it quantum mechanical} state of an 
individual molecule remain pure state and can be transformed to entangled state following the 
specific computation algorithm. 
\par The work was supported by the NIOKR RT Foundation under the grant No. 14-29.


\end{document}